\documentstyle[prd,aps]{revtex}

\begin{document}
\title{{\bf From arbitrariness to ambiguities in the evaluation of perturbative
physical amplitudes and their symmetry relations }}
\author{O.A. Battistel* and G. Dallabona**}
\maketitle

\centerline{* Dept. of Physics - CCNE, Universidade Federal de Santa Maria} 

\centerline{P.O. Box 5093, 97119-900, Santa Maria, RS, Brazil} 

\centerline{orimar@ccne.ufsm.br} 

\centerline{**Dept. of Physics - ICEx, Universidade Federal de Minas Gerais} 

\centerline{P.O. Box 702,30161-970, Belo Horizonte, MG, Brazil} 

\centerline{dalla@fisica.ufmg.br}

\begin{abstract}
A very general calculational strategy is applied to the evaluation of the
divergent physical amplitudes which are typical of perturbative
calculations. With this approach in the final results all the intrinsic
arbitrariness of the calculations due to the divergent character is still
present. We show that by using the symmetry properties as a guide to search
for the (compulsory) choices in such a way as to avoid ambiguities, a deep
and clear understanding of the role of regularization methods emerges.
Requiring then an universal point of view for the problem, as allowed by our
approach, very interesting conclusions can be stated about the possible
justifications of most intriguing aspect of the perturbative calculations in
quantum field theory: the triangle anomalies.
\end{abstract}

\vskip0.3cm

\section{Introduction}

In a certain way it seems that physicists are convinced that quantum field
theory (QFT) is the correct framework for the description of fundamental
interacting particle dynamics. Within this framework, there is a
well-established recipe for the construction of a theory for any set of
fields and symmetries previously chosen and, in principle, we can find all
the phenomenological consequences by solving the corresponding equations of
motion. However, the exact solutions, which would certainly be free from any
ambiguities, are rarely possible and we need to have recourse to
perturbative approaches in order to make predictions. As a consequence many
physical amplitudes become undefined quantities due to the presence of
divergent integrals and we have to interpret them in such a way to extract
the physical content. For this purpose it is necessary to manipulate and
calculate the divergent integrals. To make this possible, some assumptions
about the properties of such mathematical structures are required. This is
usually represented by the adoption of a regularization scheme or an
equivalent philosophy. There are many kinds of arbitrariness involved in
this step. The first one concerns the regularization technique, once this is
a choice, the final results cannot be dependent on the specific aspects
involved. The regularization technique should be only a tool to be used in
the intermediate steps. The next type of arbitrariness is related to the
routing of the internal lines. In principle, any routing should lead us to
the same physical amplitude. This is associated with the most basic symmetry
we use to construct QFT's: the space-time homogeneity. Such a property
materializes through the translational invariance of the fields, which means
in perturbative language that the amplitudes should be invariant under a
shift in the integrated momenta. Because of the divergent character,
strictly referring to cases where the degree of divergence involved is
higher than logarithmic, this is not what seems to happen. The ambiguities,
on the other hand, are frequently associated with violations of global and
local gauge symmetries. Given these aspects, manipulations and calculations
involving divergences are crucial to establish the predictive power of QFT
(in perturbative calculations).

It can be said that the discussion about divergences in the perturbative
solution of QFT has never disappeared from the literature. Recently,
however, many investigations have been focused on the questions involved in
the evaluation of perturbative amplitudes and their corresponding physical
predictions. Among others, we find the predictive power of the well-known
NJL model \cite{NJL}\cite{ORIMAR-PRD}, the implications of fermionic tensor
densities in the hadronic phenomenology of low energy {\cite{TENSOR}}, and
perhaps the most remarkable one, the controversy involved in the CPT and
Lorentz violations induced (or not) by the radiative corrections \cite{CPT1} 
\cite{CPT2}. The aim of these investigations cited above is the evaluation
of the divergent amplitudes. If in the method used some types of ambiguity
are still present, it is sufficient to prevent a definite physical
prediction. These aspects are closely related to the main motivation of the
present contribution.

The present status of this problem can be summarized in a very simple way:
in all situations where the dimensional regularization (DR) technique {\cite
{DR}} can be applied, we have at our disposal an apparently consistent
recipe to avoid violations of symmetry relations, and simultaneously the
presence of ambiguities. In situations where DR cannot be used, such as
those where the $\gamma _{5}$ matrix plays an important role, we need to
have recourse to other techniques which do not necessarily exhibit the
desirable consistency in all the problems. This is clearly not the ideal
situation and an alternative strategy should be sought so that all the
problems can be treated in a consistent way within a unique recipe. The
first requirement is that this procedure provides mappable results, one by
one, to those of DR, in situations where this technique could be applied,
but does not have any restriction of applicability out of the scope of DR.
Such a method was developed precisely with this motivation \cite{ORIMAR-TESE}%
. The strategy used consists in avoiding the explicit evaluation of
divergent integrals, using only very general properties of the eventual
regulating distribution, taken only in an implicit way in the intermediate
steps. By analyzing the final results thus obtained we can identify a set of
properties of divergent integrals that are responsible for the maintenance
of the symmetry relations. The procedure is very simple, completely
algebraic, and does not present any restriction of applicability. We will
use this procedure to state a universal point of view to analyze symmetry
relations and ambiguities related to the arbitrariness involved in the
choices of the labels for the internal line momenta. With this procedure, we
can isolate in the physical amplitudes the objects for which the role of the
chosen specific regularization becomes relevant. In a way we can say that
with our strategy we can unify such types of calculation. Precisely for this
reason a simple and rich analysis becomes possible, as will become clear in
what follows.

The purpose of this work is to investigate the possibility of treating all
the perturbative amplitudes in all theories and models, including the
anomalous ones, through a unique point of view concerning the divergences.
We will show that, as a consequence of this requirement, the ambiguities
need to be completely eliminated from the physical amplitudes, and an
alternative way to look at triangle anomalies emerges in a natural way.

We organized the work in the following way. In section II we present our
calculational strategy for manipulations and calculations of divergent
integrals. Section III is dedicated to the evaluation of some physical
amplitudes, which are analyzed in section IV, concerning the symmetry
relations and ambiguities, in order to identify the constraints on the
consistency. In section V we study the point of view of traditional
treatments for the divergences and, finally, in section VI we present our
final remarks and conclusions.

\section{Calculational Method to Manipulate Divergent Integrals}

Instead of specifying a regularization, we will adopt an alternative
strategy \cite{ORIMAR-TESE} to perform all the calculations. To justify all
the intermediate manipulation, we will assume the presence of a generic
regulating distribution only in an implicit way. This can be schematically
represented by 
\begin{equation}
\int \frac{d^{4}k}{\left( 2\pi \right) ^{4}}f(k)\rightarrow \int \frac{d^{4}k%
}{\left( 2\pi \right) ^{4}}f(k)\left\{ \lim_{\Lambda _{i}^{2}\rightarrow
\infty }G_{\Lambda _{i}}\left( k,\Lambda _{i}^{2}\right) \right\}
=\int_{\Lambda }\frac{d^{4}k}{\left( 2\pi \right) ^{4}}f(k).
\end{equation}
Here $\Lambda _{i}^{\prime }s$ are parameters of the generic distribution $%
G(\Lambda _{i}^{2},k)$ which in addition to the obvious finite character of
the modified integral must have two other very general properties. It must
be even in the integrating momentum $k$, due to Lorentz invariance
maintenance, as well as having a well-defined connection limit: i.e., 
\begin{equation}
\lim_{\Lambda _{i}^{2}\rightarrow \infty }G_{\Lambda _{i}}\left(
k^{2},\Lambda _{i}^{2}\right) =1.
\end{equation}
The first property implies that all odd integrands vanish, while the second
one guarantees, in particular, that the value of the finite integrals in the
amplitudes will not be modified. Note that these requirements are completely
general and are in agreement with any reasonable 4D regularization. After
these assumptions we can manipulate the integrand of the divergent integrals
by using identities to generate a mathematical expression where all the
divergences are contained in momentum independent structures. Due to the
fact that in perturbative amplitudes we always have propagators, an adequate
identity to achieve this goal is the following \cite{ORIMAR-TESE} 
\begin{equation}
\frac{1}{[(k+k_{i})^{2}-m^{2}]}=\sum_{j=0}^{N}\frac{\left( -1\right)
^{j}\left( k_{i}^{2}+2k_{i}\cdot k\right) ^{j}}{\left( k^{2}-m^{2}\right)
^{j+1}}+\frac{\left( -1\right) ^{N+1}\left( k_{i}^{2}+2k_{i}\cdot k\right)
^{N+1}}{\left( k^{2}-m^{2}\right) ^{N+1}\left[ \left( k+k_{i}\right)
^{2}-m^{2}\right] },
\end{equation}
where $k_{i}$ is (in principle) an arbitrary momentum used in the routing of
an internal line. The value for $N$ in the above expression can be
adequately chosen to avoid unnecessary algebraic difficulty. It can be taken
as the smallest value that leads the last term in the above expression to a
finite integral. As a consequence, all the momentum dependent parts of the
amplitudes can be integrated without restrictions due to the connection
limit requirement. The divergent structures obtained in this way, on the
other hand, have no additional assumptions, and (in the present discussion)
they are written as a combination of five objects: namely, 
\begin{eqnarray}
\bullet \Box _{\alpha \beta \mu \nu } &=&\int_{\Lambda }\frac{d^{4}k}{\left(
2\pi \right) ^{4}}\frac{24k_{\mu }k_{\nu }k_{\alpha }k_{\beta }}{\left(
k^{2}-m^{2}\right) ^{4}}-g_{\alpha \beta }\int_{\Lambda }\frac{d^{4}k}{%
\left( 2\pi \right) ^{4}}\frac{4k_{\mu }k_{\nu }}{\left( k^{2}-m^{2}\right)
^{3}}  \nonumber \\
&&-g_{\alpha \nu }\int_{\Lambda }\frac{d^{4}k}{\left( 2\pi \right) ^{4}}%
\frac{4k_{\beta }k_{\mu }}{\left( k^{2}-m^{2}\right) ^{3}}-g_{\alpha \mu
}\int_{\Lambda }\frac{d^{4}k}{\left( 2\pi \right) ^{4}}\frac{4k_{\beta
}k_{\nu }}{\left( k^{2}-m^{2}\right) ^{3}}, \\
\bullet \Delta _{\mu \nu } &=&\int_{\Lambda }\frac{d^{4}k}{\left( 2\pi
\right) ^{4}}\frac{4k_{\mu }k_{\nu }}{\left( k^{2}-m^{2}\right) ^{3}}%
-\int_{\Lambda }\frac{d^{4}k}{\left( 2\pi \right) ^{4}}\frac{g_{\mu \nu }}{%
\left( k^{2}-m^{2}\right) ^{2}}, \\
\bullet \nabla _{\mu \nu } &=&\int_{\Lambda }\frac{d^{4}k}{\left( 2\pi
\right) ^{4}}\frac{2k_{\nu }k_{\mu }}{\left( k^{2}-m^{2}\right) ^{2}}%
-\int_{\Lambda }\frac{d^{4}k}{\left( 2\pi \right) ^{4}}\frac{g_{\mu \nu }}{%
\left( k^{2}-m^{2}\right) }, \\
\bullet I_{log}(m^{2}) &=&\int_{\Lambda }\frac{d^{4}k}{\left( 2\pi \right)
^{4}}\frac{1}{\left( k^{2}-m^{2}\right) ^{2}}, \\
\bullet I_{quad}(m^{2}) &=&\int_{\Lambda }\frac{d^{4}k}{\left( 2\pi \right)
^{4}}\frac{1}{\left( k^{2}-m^{2}\right) }.
\end{eqnarray}
It is important to emphasize that with this strategy it becomes possible to
map the final expressions obtained by us onto the corresponding results from
other techniques, due to the fact that all the steps are perfectly valid
within reasonable regularization prescriptions, including the DR technique.
All we need is to evaluate the divergent structures obtained in the specific
philosophy that we want to contact. In addition {\it we focus on the fact \
that no shifts will be used on the general routing assumed for all \
amplitudes. Consequently, it will be possible to contact those results \
corresponding to the explicit evaluation of surface terms in the eventual
shifts \ performed in the integrating momentum of the loop integrals}. This
very general character of the adopted strategy will become the most
important ingredient for the analysis we want to do, and to support our
conclusions. Another important aspect of the procedure is that a definite
value is attributed to each divergent integral involved. This value is used
every time the integrals are present in a physical amplitude in all theories
and models, providing in this way a universal point of view for the problem.
No expansions, limits, or not totally controlled procedures are included.
All the manipulations and calculations we need in the treatment of
divergences in QFT are performed without the explicit calculation of a
divergent integral, as will be explained.

Another aspect we want to focus on is the question of the intrinsic
arbitrariness of the perturbative calculations implied by the divergences.
All the arbitrariness is still present in our final expressions. The
arbitrariness involved in the choice of labels for the internal lines is
maintained by taking the most general ones and not performing shifts in the
intermediate steps. The arbitrariness involved in the choice of the value to
be assumed for the undefined mathematical objects (choice of regularization)
is also present. We can say that with our strategy we go as far as possible
in the manipulations and calculations of the divergent amplitudes. The next
step after our calculations necessarily involves a arbitrariness. The
advantage of our procedure is precisely to allow us to make a clear and
transparent analysis of the problem. The conclusions may help us to get
understanding about some intriguing questions associated with the
divergences in perturbative calculations; in particular, the question of the
ambiguities, that is, the possibility that the physical amplitudes become
dependent on the  arbitrariness involved. We are naturally guided by the
following desire: to choose what must be chosen, but to do it in a way so
that the physical consequences do not become dependent on our choices, as we
have always learned in physics since our first lessons.

In order to clarify the procedure described above, let us explicitly
evaluate some simple but representative divergent integrals, defined as 
\begin{equation}
\left( I_{1};I_{1}^{\mu }\right) =\int \frac{d^{4}k}{(2\pi )^{4}}\frac{%
\left( 1;k^{\mu }\right) }{[(k+k_{1})^{2}-m^{2}]}.
\end{equation}
The first one to be treated may be the one which corresponds to a higher
degree of divergence in fundamental theories: the cubic one. As a first
step, after assuming the presence of a generic distribution on the
integrand, we write 
\begin{eqnarray}
\left( I_{1}\right) _{\mu } &=&-\int_{\Lambda }\frac{d^{4}k}{(2\pi )^{4}}%
\frac{2k_{\mu }\left( k_{1}\cdot k\right) }{(k^{2}-m^{2})^{2}}+k_{1\nu
}k_{1\alpha }k_{1\beta }\left\{ \int_{\Lambda }\frac{d^{4}k}{(2\pi )^{4}}%
\frac{4g_{\alpha \beta }k_{\mu }k_{\nu }}{(k^{2}-m^{2})^{3}}-\int_{\Lambda }%
\frac{d^{4}k}{(2\pi )^{4}}\frac{8k_{\alpha }k_{\beta }k_{\mu }k_{\nu }}{%
(k^{2}-m^{2})^{4}}\right\}   \nonumber \\
&&-\left\{ \int \frac{d^{4}k}{(2\pi )^{4}}\frac{6k_{1}^{4}\left( k_{1}\cdot
k\right) k_{\mu }}{(k^{2}-m^{2})^{4}}-\int \frac{d^{4}k}{(2\pi )^{4}}\frac{%
\left( k_{1}^{2}+2k_{1}\cdot k\right) ^{4}k_{\mu }}{%
(k^{2}-m^{2})^{4}[(k+k_{1})^{2}-m^{2}]}\right\} .
\end{eqnarray}
On the right hand side of the above equation, we used the identity (3) with
the choice $N=3$, ruled out an odd integral, and dropped the subscript $%
\Lambda $ in the last two integrals, due to their finite character
(connection limit requirement). Now we can perform the integration of the
finite terms without any restriction. Using standard techniques to solve
finite Feynman integrals we can easily verify that an exact cancellation is
obtained. Then we rewrite the result in terms of the objects (4)-(8) 
\begin{equation}
\left( I_{1}\right) _{\mu }=-k_{1\mu }I_{quad}(m^{2})-k_{1\beta }\left[
\nabla _{\beta \mu }\right] -\frac{1}{3}k_{1\beta }k_{1\alpha }k_{1\nu }%
\left[ \Box _{\alpha \beta \mu \nu }\right] -\frac{1}{3}k_{1\mu }k_{1\alpha
}k_{1\beta }\left[ \triangle _{\alpha \beta }\right] +\frac{1}{3}%
k_{1}^{2}k_{1\nu }\left[ \triangle _{\mu \nu }\right] .
\end{equation}
Note that, as was said before, no specific assumption about the divergent
integrals was made, which means that for these steps all reasonable
regularizations could be used. This is equivalent to applying the
regularization on the obtained expression eq.(11). Following a similar
procedure we also obtain 
\begin{equation}
I_{1}=I_{quad}(m^{2})+k_{1\mu }k_{1\nu }\left[ \Delta _{\mu \nu }\right] .
\end{equation}

Others typical divergent integrals we need to use in perturbative
calculations are those constructed with two propagators, defined as 
\begin{equation}
\left( I_{2};I_{2}^{\mu };I_{2}^{\mu \nu }\right) =\int \frac{d^{4}k}{(2\pi
)^{4}}\frac{\left( 1;k^{\mu };k^{\mu }k^{\nu }\right) }{%
[(k+k_{1})^{2}-m^{2}][(k+k_{2})^{2}-m^{2}]}.
\end{equation}
Let us take the linearly divergent one, which is the following: 
\begin{eqnarray}
\left( I_{2}\right) _{\mu } &=&-\frac{1}{2}(k_{1}+k_{2})_{\alpha
}\int_{\Lambda }\frac{d^{4}k}{(2\pi )^{4}}\frac{4k_{\alpha }k_{\mu }}{%
(k^{2}-m^{2})^{3}}+\int \frac{d^{4}k}{(2\pi )^{4}}\frac{(k_{1}^{2}+2k_{1}%
\cdot k)^{2}k_{\mu }}{(k^{2}-m^{2})^{3}\left[ (k+k_{1})^{2}-m^{2}\right] } 
\nonumber \\
&&+\int \frac{d^{4}k}{(2\pi )^{4}}\frac{(k_{2}^{2}+2k_{2}\cdot k)^{2}k_{\mu }%
}{(k^{2}-m^{2})^{3}\left[ (k+k_{2})^{2}-m^{2}\right] }+\int \frac{d^{4}k}{%
(2\pi )^{4}}\frac{(k_{1}^{2}+2k_{1}\cdot k)(k_{2}^{2}+2k_{2}\cdot k)k_{\mu }%
}{(k^{2}-m^{2})^{2}\left[ (k+k_{1})^{2}-m^{2}\right] \left[
(k+k_{2})^{2}-m^{2}\right] }.
\end{eqnarray}
In this case we used the identity (3) with $N=2$ to rewrite both
propagators, and again we ruled out an odd integral and dropped the
subscript $\Lambda $ on the finite integrals thus obtained. Solving then the
finite ones, we obtain 
\begin{equation}
\left( I_{2}\right) _{\mu }=-\frac{1}{2}(k_{1}+k_{2})_{\alpha }\left[ \Delta
_{\alpha \mu }\right] -\frac{1}{2}(k_{1}+k_{2})_{\mu }\left\{ I_{\log
}(m^{2})-\left( \frac{i}{(4\pi )^{2}}\right)
Z_{0}(m^{2},m^{2},(k_{1}-k_{2})^{2};m^{2})\right\} .
\end{equation}
Here we introduced the one-loop structure functions defined by 
\begin{equation}
Z_{k}(\lambda _{1}^{2},\lambda _{2}^{2},q^{2};\lambda
^{2})=\int_{0}^{1}dzz^{k}ln\left( \frac{q^{2}z(1-z)+(\lambda
_{1}^{2}-\lambda _{2}^{2})z-\lambda _{1}^{2}}{(-\lambda ^{2})}\right) ,
\end{equation}
which can be explicitly solved, but for the present discussion this is
immaterial. In order to simplify the notation, from now on we will adopt $%
Z_{0}(m^{2},m^{2},q^{2};m^{2})=Z_{0}(q^{2};m^{2})$ since we are dealing with
only one species of intermediate fermion.

Following strictly the same steps we can obtain the result 
\begin{equation}
I_{2}=I_{log}(m^{2})-\left( \frac{i}{(4\pi )^{2}}\right)
Z_{0}((k_{1}-k_{2})^{2};m^{2}),
\end{equation}
and after some algebraic effort 
\begin{eqnarray}
&&\left( I_{2}\right) _{\mu \nu }=\int_{\Lambda }\frac{d^{4}k}{(2\pi )^{4}}%
\frac{k_{\mu }k_{\nu }}{(k^{2}-m^{2})^{2}}  \nonumber \\
&&\;\;\;\;\;\;\;\;\;\;-\left\{ (k_{1}^{2}+k_{2}^{2})\int_{\Lambda }\frac{%
d^{4}k}{(2\pi )^{4}}\frac{k_{\mu }k_{\nu }}{(k^{2}-m^{2})^{3}}+(k_{1\alpha
}k_{1\beta }+k_{2\alpha }k_{2\beta }+k_{1\beta }k_{2\alpha })\int_{\Lambda }%
\frac{d^{4}k}{(2\pi )^{4}}\frac{4k_{\alpha }k_{\beta }k_{\mu }k_{\nu }}{%
(k^{2}-m^{2})^{4}}\right\}   \nonumber \\
&&\;\;\;\;\;\;\;\;\;\;+\left. \left( \frac{i}{(4\pi )^{2}}\right) \right\{ %
\left[ (k_{1}-k_{2})_{\mu }(k_{1}-k_{2})_{\nu }-(k_{1}-k_{2})^{2}g_{\mu \nu }%
\right] \left[ -Z_{2}((k_{1}-k_{2})^{2};m^{2})+\frac{1}{4}%
Z_{0}((k_{1}-k_{2})^{2};m^{2})\right]   \nonumber \\
&&\;\;\;\;\;\;\;\;\;\;\;\;\;\;\;\;\;\;\;\;\;\;\;\;\;\;\;\;\;\;\left. -\frac{1%
}{4}(k_{1}+k_{2})_{\mu }(k_{1}+k_{2})_{\nu
}Z_{0}((k_{1}-k_{2})^{2};m^{2})\right\} .
\end{eqnarray}
It is easy to note that all Feynman integrals can be easily treated by the
described procedure. It is always possible to identify a set of structure
functions analogous to the $Z_{k}$ ones, irrespective of the number of
propagators and loops involved. If there are more loops, new basic objects
will be added to the set (4)-(8). For the present discussion the considered
set of integrals will be enough.

\section{Physical Amplitudes}

In this section we consider some specific examples of amplitudes that can be
evaluated using the divergent integrals evaluated in the previous section.
Although a few cases will be considered, the conclusions become transparent
concerning the aspects we want to focus on. The first simple and important
case is the one-point vector function related, for example, to the tadpole
diagram in the lowest order of the electron self-energy in quantum
electrodynamics (QED). It is defined by 
\begin{equation}
T_{\mu }^{V}(k_{1},m)=\int \frac{d^{4}k}{(2\pi )^{4}}Tr\left\{ \gamma _{\mu }%
\frac{1}{(\not{k}+\not{k}_{1})-m}\right\} ,
\end{equation}
where $k_{1}$ represents the arbitrary choice for the routing of internal
line momentum and $m$ is the $\frac{1}{2}$ fermion mass. After the Dirac
trace evaluation we get 
\begin{equation}
T_{\mu }^{V}=4\left\{ \int \frac{d^{4}k}{(2\pi )^{4}}\frac{k_{\mu }}{%
(k+k_{1})^{2}-m^{2}}+k_{1\mu }\int \frac{d^{4}k}{(2\pi )^{4}}\frac{1}{%
(k+k_{1})^{2}-m^{2}}\right\} .
\end{equation}
Inserting then the results (11) and (12), we get 
\begin{equation}
T_{\mu }^{V}=4\left\{ -k_{1\beta }\left[ \nabla _{\beta \mu }\right] -\frac{1%
}{3}k_{1\beta }k_{1\alpha }k_{1\nu }\left[ \Box _{\alpha \beta \mu \nu }%
\right] +\frac{1}{3}k_{1}^{2}k_{1\nu }\left[ \triangle _{\nu \mu }\right] +%
\frac{2}{3}k_{1\mu }k_{1\alpha }k_{1\beta }\left[ \triangle _{\alpha \beta }%
\right] \right\} .
\end{equation}
A similar structure is the scalar one-point function present in the fermion
self-energy when, in the theory, there is a coupling with a scalar field,
for which only the result (12) is necessary: 
\begin{equation}
T^{S}(k_{1},m)=4m\left\{ I_{quad}(m^{2})+k_{1}^{\beta }k_{1}^{\alpha }\left[
\triangle _{\beta \alpha }\right] \right\} .
\end{equation}
Next, with the integrals (15), (17), and (18) we can construct a set of
two-point functions playing an important role in lowest order contributions
for the self-energies of intermediate bosons. They can be defined by the
expression 
\begin{equation}
T^{ij}(k_{1},k_{2};m)=\int \frac{d^{4}k}{(2\pi )^{4}}Tr\left\{ \Gamma _{i}%
\frac{1}{\not{k}+{\not{k}}_{1}-m}\Gamma _{j}\frac{1}{\not{k}+{\not{k}}_{2}-m}%
\right\} .
\end{equation}
Here $k_{1}$ and $k_{2}$ represent arbitrary choices for the internal line
momenta. The difference between them is the external momentum. Taking the
operators $\Gamma _{i}=[\hat{1};\gamma _{5};\gamma _{\mu };i\gamma _{\mu
}\gamma _{5}]=(S,P,V,A),$ a set of functions can be constructed. The
advantage of the use of these functions in our discussion is that they are
important structures in fundamental theories and that they have very simple
symmetry relations among them, which work as necessary conditions for all
calculational methods to get the desired consistency in the divergent
aspects involved. The first two-point function we take into account is the
simplest one $T_{\mu }^{AP},$ written with the help of eq.(17) as 
\begin{equation}
T_{\mu }^{AP}=-4mi(k_{1}-k_{2})_{\mu }\left\{ I_{log}(m^{2})-\left( \frac{i}{%
(4\pi )^{2}}\right) Z_{0}((k_{1}-k_{2})^{2};m^{2})\right\} .
\end{equation}
From this, with the results (15) and (17) we can write down the expression
\begin{equation}
T_{\mu }^{VS}=-4m(k_{1}+k_{2})_{\beta }[\triangle _{\beta \mu }],
\end{equation}
and the very important one for our future analysis 
\begin{equation}
T_{\mu \nu }^{AV}=-2\varepsilon _{\mu \nu \alpha \beta }(k_{1}-k_{2})_{\beta
}(k_{1}+k_{2})_{\xi }[\triangle _{\xi \alpha }].
\end{equation}
The results (12) and (17) allows us to write 
\begin{eqnarray}
T^{SS} &=&4\left\{ I_{quad}(m^{2})+\frac{4m^{2}-(k_{1}-k_{2})^{2}}{2}%
I_{log}(m^{2})-\frac{4m^{2}-(k_{1}-k_{2})^{2}}{2}\left( \frac{i}{(4\pi )^{2}}%
\right) Z_{0}((k_{1}-k_{2})^{2};m^{2})\right\}   \nonumber \\
&&+(k_{1}-k_{2})_{\alpha }(k_{1}-k_{2})_{\beta }\left[ \triangle _{\alpha
\beta }\right] +(k_{1}+k_{2})_{\alpha }(k_{1}+k_{2})_{\beta }\left[
\triangle _{\alpha \beta }\right] ,
\end{eqnarray}
and 
\begin{eqnarray}
T^{PP} &=&4\left\{ -I_{quad}(m^{2})+\frac{1}{2}%
(k_{1}-k_{2})^{2}I_{log}(m^{2})-\frac{1}{2}(k_{1}-k_{2})^{2}\left( \frac{i}{%
(4\pi )^{2}}\right) Z_{0}((k_{1}-k_{2})^{2};m^{2})\right\}   \nonumber \\
&&-(k_{1}-k_{2})_{\alpha }(k_{1}-k_{2})_{\beta }\left[ \triangle _{\alpha
\beta }\right] -(k_{1}+k_{2})_{\alpha }(k_{1}+k_{2})_{\beta }\left[
\triangle _{\alpha \beta }\right] .
\end{eqnarray}
Finally the results (12),(15),(17), and (18) lead us to the expressions 
\begin{eqnarray}
T_{\mu \nu }^{VV} &=&\frac{4}{3}[(k_{1}-k_{2})^{2}g_{\mu \nu
}-(k_{1}-k_{2})_{\mu }(k_{1}-k_{2})_{\nu }]\times   \nonumber \\
&&\times \left\{ I_{log}(m^{2})-\left( \frac{i}{(4\pi )^{2}}\right) \left[ 
\frac{1}{3}+\frac{2m^{2}+(k_{1}-k_{2})^{2}}{(k_{1}-k_{2})^{2}}%
Z_{0}((k_{1}-k_{2})^{2};m^{2})\right] \right\} +A_{\mu \nu },
\end{eqnarray}
and 
\begin{eqnarray}
T_{\mu \nu }^{AA} &=&-\frac{4}{3}[(k_{1}-k_{2})^{2}g_{\mu \nu
}-(k_{1}-k_{2})_{\mu }(k_{1}-k_{2})_{\nu }]\left\{ I_{log}(m^{2})-\left( 
\frac{i}{(4\pi )^{2}}\right) \left[ \frac{1}{3}+\frac{%
2m^{2}+(k_{1}-k_{2})^{2}}{(k_{1}-k_{2})^{2}}Z_{0}((k_{1}-k_{2})^{2};m^{2})%
\right] \right\}   \nonumber \\
&&+g_{\mu \nu }8m^{2}\left\{ I_{log}(m^{2})-\left( \frac{i}{(4\pi )^{2}}%
\right) Z_{0}((k_{1}-k_{2})^{2};m^{2})\right\} -A_{\mu \nu },
\end{eqnarray}
where 
\begin{eqnarray}
A_{\mu \nu } &=&4[\nabla _{\mu \nu }]+(k_{1}-k_{2})_{\alpha
}(k_{1}-k_{2})_{\beta }\left[ \frac{1}{3}\Box _{\alpha \beta \mu \nu }+\frac{%
1}{3}\triangle _{\mu \beta }g_{\alpha \nu }+g_{\alpha \mu }\triangle _{\beta
\nu }-g_{\mu \nu }\triangle _{\alpha \beta }-\frac{2}{3}g_{\alpha \beta
}\triangle _{\mu \nu }\right]   \nonumber \\
&&+\left[ (k_{1}-k_{2})_{\alpha }(k_{1}+k_{2})_{\beta
}-(k_{1}+k_{2})_{\alpha }(k_{1}-k_{2})_{\beta }\right] \left[ \frac{1}{3}%
\Box _{\alpha \beta \mu \nu }+\frac{1}{3}\triangle _{\mu \beta }g_{\nu
\alpha }+\frac{1}{3}\triangle _{\beta \nu }g_{\alpha \mu }\right]   \nonumber
\\
&&+(k_{1}+k_{2})_{\alpha }(k_{1}+k_{2})_{\beta }\left[ \Box _{\alpha \beta
\mu \nu }-\triangle _{\nu \alpha }g_{\mu \beta }-\triangle _{\beta \nu
}g_{\alpha \mu }-3\triangle _{\alpha \beta }g_{\mu \nu }\right] .
\end{eqnarray}
It is important to focus on very general aspects of the result obtained.
First, all the steps can be easily identified as valid in any regularization
scheme, since no specific calculations for the divergent integrals were
made. Only safe steps were performed. The results were presented in terms of
a minimum number of structures, which allows us to make a very simple and
clear analysis. The choices for the routing of the internal line momenta
were taken as general as possible. Consequently, all the potentially
ambiguous terms still remain in the results. Let us now consider the
constraints imposed for symmetry reasons in the explicit expressions of the
perturbative physical amplitudes.

\section{Symmetry Relations}

It is well known that the symmetry content of a QFT, implemented in the
construction of the invariant Lagrangian, states specific relations for the
perturbative amplitudes in all orders. We refer to the Ward-Takarashi and
the Slanov-Taylor identities, and other general constraints like Furry's
theorem. This implies that the calculated amplitudes, independent of the
specific aspects of the regularization technique applied, must obey these
relations. It is very desirable that such properties in all theories and
models should be simultaneously preserved, where they must be, as a
consequence of a unique rule, and that the correct result for the violations
should emerge in a natural way without changing the adopted rules, as is the
case with the triangle anomalies. We understand that with our approach some
clarifications about these aspects can be given since there is no
restriction of applicability. Through the adopted strategy it becomes
possible to seek the necessary properties so that the divergent integrals
lead us to the desired consistency. After that, if one wants, one can
construct a consistent technique. This means inverting the usual procedure,
which is to propose at the starting point a specific regularization, and
then to test the consistency. Having this in mind, we can verify a set of
Ward identities relating the fermionic Green's functions directly dictated
by the vector current conservation and by the proportionality between the
axial-vector and pseudoscalar currents, as well as other constraints imposed
by CPT symmetries.

We can start with the simplest amplitude that carries a Lorentz index, the $%
T_{\mu }^{V}(l,m)$. The general symmetry grounds in Furry's theorem state
that only a vanishing value is reasonable, so we need to satisfy 
\begin{equation}
0=-k_{1\beta }\nabla _{\beta \mu }-\frac{1}{3}k_{1\beta }k_{1\alpha }k_{1\nu
}\Box _{\alpha \beta \mu \nu }+\frac{1}{3}k_{1}^{2}k_{1\nu }\triangle _{\nu
\mu }+\frac{2}{3}k_{1\mu }k_{1\alpha }k_{1\beta }\triangle _{\alpha \beta }.
\end{equation}
There are two different ways to satisfy the above result. The first one is
the choice $k_{1}=0$ since $k_{1}$ is arbitrary. We can question if it is
always possible to make this choice. Thinking separately about this
amplitude, the answer can be positive, but we note that it is possible to
relate this amplitude to other purely fermionic vector functions which have
more than one Lorentz index, like $T_{\mu \nu }^{VV}$. This relation can
easily be constructed by noting the identity 
\begin{equation}
(k_{1}-k_{2})_{\mu }\left\{ \gamma _{\nu }\frac{1}{(\not{k}+{\not{k}}_{1})-m}%
\gamma _{\mu }\frac{1}{(\not{k}+{\not{k}}_{2})-m}\right\} =\gamma _{\nu }%
\frac{1}{[\not{k}+\not{k}_{2}-m]}-\gamma _{\nu }\frac{1}{[\not{k}+\not%
{k}_{1}-m]}.
\end{equation}
Taken the traces and integrating both sides in the momentum $k$ we get the
relation 
\begin{equation}
(k_{1}-k_{2})_{\mu }T_{\mu \nu }^{VV}=T_{\nu }^{V}(k_{2};m)-T_{\nu
}^{V}(k_{1};m),
\end{equation}
which also needs to be zero due to vector current conservation, since $k_{1}$
and $k_{2}$ refer to momenta of the $T_{\mu \nu }^{VV}$ amplitude. The
choice $k_{1}=k_{2}=0$ simultaneously cannot be at our disposal, so we need
to use a second way to satisfy eq.(32). We construct a regularization
requiring that the properties 
\begin{equation}
\Box _{\alpha \beta \mu \nu }^{reg}=\nabla _{\mu \nu }^{reg}=\triangle _{\mu
\nu }^{reg}=0,
\end{equation}
are included, which from now on we denominate {\it consistency conditions. }%
The same conclusion can be extracted if we consider the explicit expression
for $T_{\mu \nu }^{VV}$, eq.(29), and contract it with the external momentum 
\begin{eqnarray}
(k_{1}-k_{2})_{\mu }T_{\mu \nu }^{VV} &=&4\left\{ (k_{1}-k_{2})_{\alpha
}\nabla _{\alpha \nu }+(k_{1\alpha }k_{1\beta }k_{1\rho }-k_{2\alpha
}k_{2\beta }k_{2\rho })\frac{\Box _{\alpha \beta \rho \nu }}{3}\right.  
\nonumber \\
&&\;\;\;\;\;\left. -(k_{1}^{2}k_{1\rho }-k_{2}^{2}k_{2\rho })\frac{\triangle
_{\rho \nu }}{3}-(k_{1\nu }k_{1\alpha }k_{1\beta }-k_{2\nu }k_{2\alpha
}k_{2\beta })\frac{2}{3}\triangle _{\alpha \beta }\right\} .
\end{eqnarray}
Note that a conserved vector current cannot be obtained by thinking about
convenient choices for the arbitrary momenta $k_{1}$ and $k_{2}$, due to the
fact that there are violating terms, coefficients of all pieces $\Box
_{\alpha \beta \mu \nu },\nabla _{\mu \nu },$ and $\triangle _{\mu \nu },$
which are independent of such eventual choices (nonambiguous). It seems
there is no way to escape the conditions (35). In a similar way to $T_{\mu
}^{V}$, some considered two-point functions need to vanish identically on
general symmetry grounds. They are $T_{\mu }^{VS}$ and $T_{\mu \nu }^{AV}$
which state respectively 
\begin{equation}
\left\{ 
\begin{array}{l}
(k_{1}+k_{2})_{\beta }\triangle _{\mu \beta }=0 \\ 
\varepsilon _{\mu \nu \alpha \beta }(k_{2}-k_{1})_{\beta }(k_{1}+k_{2})_{\xi
}\triangle _{\xi \alpha }=0.
\end{array}
\right. 
\end{equation}
For these two specific situations, both options are available: choosing $%
k_{1}$ and $k_{2}$ in a convenient way and constructing $\triangle _{\mu
\beta }^{reg}=0.$ It is also interesting to consider the contractions with
external momenta once symmetry relations can be generated. For such
contractions we get 
\begin{eqnarray}
\bullet (k_{1}-k_{2})_{\mu }T_{\mu }^{VS} &=&-4m(k_{1}-k_{2})_{\mu
}(k_{1}+k_{2})_{\beta }\triangle _{\mu \beta } \\
\bullet (k_{1}-k_{2})_{\mu }T_{\mu \nu }^{AV} &=&2\varepsilon _{\mu \nu
\alpha \beta }(k_{1}-k_{2})_{\mu }(k_{2}-k_{1})_{\beta }(k_{1}+k_{2})_{\xi
}\triangle _{\xi \alpha } \\
\bullet (k_{1}-k_{2})_{\nu }T_{\mu \nu }^{AV} &=&2\varepsilon _{\mu \nu
\alpha \beta }(k_{1}-k_{2})_{\nu }(k_{2}-k_{1})_{\beta }(k_{1}+k_{2})_{\xi
}\triangle _{\xi \alpha }.
\end{eqnarray}
A conserved vector current for $T_{\mu }^{VS}$ can be obtained by the choice 
$k_{1}=-k_{2}$, as well as by taking $\triangle _{\mu \nu }^{reg}=0.$
However, both contractions involving $T_{\mu \nu }^{AV}$ vanish identically
independent of the possible choices, just because the antisymmetric $%
\varepsilon _{\mu \nu \alpha \beta }$ is contracted with a symmetric object
in two of its indices. The vector current must be conserved, but the
axial-vector current must not. So there is only one consistent value for $%
T_{\mu \nu }^{AV}$: the identically zero value. Otherwise a symmetry
relation is broken. We can add to this argumentation another very general
aspect that forces us to obtain a zero value for $T_{\mu \nu }^{AV}$ (and $%
T_{\mu }^{VS}$): unitarity. If the amplitude does not vanish then it needs
to develop an imaginary part at $(k_{1}-k_{2})^{2}=4m^{2}$ to be consistent
with unitarity (Cutkosky's rules). Clearly, independent of the possibilities
involved, this property cannot be present in the results (25) and (26) for $%
T_{\mu }^{VS}$ and $T_{\mu \nu }^{AV},$ respectively. The above
argumentation will become very important.

Now we can take the $T_{\mu \nu }^{AA}$ symmetry content for analysis. The
proportionality between the axial-vector and the pseudoscalar current states
that we need to obtain is $(k_{1}-k_{2})_{\mu }T_{\mu \nu }^{AA}=-2miT_{\nu
}^{PA}$, and another similar condition for the contraction involving the $%
\nu $ Lorentz index. Using the explicit expression obtained for $T_{\mu \nu
}^{AA}$, eq.(30), we get 
\begin{eqnarray}
(k_{1}-k_{2})_{\mu }T_{\mu \nu }^{AA} &=&4\left\{ (k_{2}-k_{1})_{\alpha
}\nabla _{\alpha \nu }+(k_{1\alpha }k_{1\beta }k_{1\rho }-k_{2\alpha
}k_{2\beta }k_{2\rho })\frac{\Box _{\alpha \beta \rho \nu }}{3}\right.  
\nonumber \\
&&\;\;\;\;\;\;\left. +(k_{1}^{2}k_{1\rho }-k_{2}^{2}k_{2\rho })\frac{%
\triangle _{\rho \nu }}{3}+(k_{1\nu }k_{1\alpha }k_{1\beta }-k_{2\nu
}k_{2\alpha }k_{2\beta })\frac{2}{3}\triangle _{\alpha \beta }\right\}
-2miT_{\nu }^{PA}.
\end{eqnarray}
The constraints are completely similar to those obtained for $T_{\mu \nu
}^{VV}$, since $T_{\nu }^{PA}$ is nonambiguous. These conditions can be put
in terms of the value for the $T_{\nu }^{V}$ amplitude, irrespective of the $%
T_{\nu }^{PA}$ calculation, if we note the identity 
\begin{eqnarray}
(k_{1}-k_{2})_{\mu }\left\{ i\gamma _{\nu }\gamma _{5}\frac{1}{\not{k}+{\not%
{k}}_{1}-m}i\gamma _{\mu }\gamma _{5}\frac{1}{\not{k}+{\not{k}}_{2}-m}%
\right\}  &=&2mi\left\{ i\gamma _{\nu }\gamma _{5}\frac{1}{\not{k}+{\not{k}}%
_{1}-m}\gamma _{5}\frac{1}{\not{k}+{\not{k}}_{2}-m}\right\}   \nonumber \\
&&-\gamma _{\nu }\frac{1}{\not{k}+\not{k}_{2}-m}-\gamma _{\nu }\gamma _{5}%
\frac{1}{\not{k}+\not{k}_{1}-m}\gamma _{5}.
\end{eqnarray}
After taking the traces and integrating both sides in the momentum $k,$\
eq.(41) results. The above procedure can be used to relate the conditions to
be imposed on the $(n-1)$-point Green's functions to the Ward identity
involving $n$-point Green's functions without explicit evaluation of the $n$%
-point structures. Frequently, this is a procedure used in discussions
relative to violation of symmetry relations or the triangle anomaly
phenomenon. This is a very important aspect of perturbative QFT. Let us then
investigate the Ward identities involving three-point functions in this way.
We can define all of them as 
\begin{equation}
T^{ijk}=\int \frac{d^{4}k}{(2\pi )^{4}}Tr\left\{ \Gamma _{i}\frac{1}{\not{k}+%
{\not{k}}_{1}-m}\Gamma _{j}\frac{1}{\not{k}+{\not{k}}_{2}-m}\Gamma _{k}\frac{%
1}{\not{k}+{\not{k}}_{3}-m}\right\} .
\end{equation}
Here the $\Gamma $ operators assume $[\hat{1};\gamma _{5};\gamma _{\mu
};i\gamma _{\mu }\gamma _{5}]=(S,P,V,A)$ and $k_{1},k_{2},k_{3}$ represent
arbitrary choices for the involved momenta of the internal lines. Only their
differences have a physical meaning. A physical process requires the
symmetrization of the final states. So to construct a symmetry relation we
need to include the direct and the crossed diagrams.

In order to clarify the notation, let us consider an example. If $\Gamma
_{i}=\gamma _{\mu },\Gamma _{j}=\gamma _{\nu }$ and $\Gamma _{k}=1$, the
physical process, symmetrized in the final states, is written as 
\begin{equation}
T_{\mu \nu }^{S\rightarrow VV}=T_{\mu \nu
}^{SVV}(k_{1},k_{2},k_{3};m)+T_{\nu \mu }^{SVV}(l_{1},l_{2},l_{3};m),
\end{equation}
where the first term means the direct channel with external momenta defined
by $k_{3}-k_{2}=q,k_{3}-k_{1}=p,k_{1}-k_{2}=p^{\prime }$, and the second one
is the crossed channel where $l_{3}-l_{2}=q,l_{3}-l_{1}=p^{\prime
},l_{1}-l_{2}=p$. Note that the routing is taken as the most general one,
diagram by diagram, since the eventual ambiguities present in a particular
Green's function do not have the same meaning as those associated with any
other, even though the external momenta are the same ones. If we take the
same meaning for the parametrization of the routing of internal lines in
both diagrams we will be assuming a universal meaning for the completely
undefined quantities present in a Feynman diagram.

After these necessary definitions we consider in a more detailed way a very
important case of such a three-point function: $T_{\lambda \mu \nu }^{AVV}$.
To generate a relation among these Green's functions and others, we first
note the convenient identity 
\begin{eqnarray}
&&(k_{3}-k_{2})_{\lambda }\left\{ \gamma _{\nu }\frac{1}{\not{k}+{\not{k}}%
_{2}-m}i\gamma _{\lambda }\gamma _{5}\frac{1}{\not{k}+{\not{k}}_{3}-m}\gamma
_{\mu }\frac{1}{\not{k}+{\not{k}}_{1}-m}\right\} =-\left\{ i\gamma _{\nu
}\gamma _{5}\frac{1}{\not{k}+{\not{k}}_{3}-m}\gamma _{\mu }\frac{1}{\not{k}+{%
\ \not{k}}_{1}-m}\right\}   \nonumber \\
&&+\left\{ \gamma _{\nu }\frac{1}{\not{k}+{\not{k}}_{2}-m}i\gamma _{\mu
}\gamma _{5}\frac{1}{\not{k}+{\not{k}}_{1}-m}\right\} -2mi\left\{ \gamma
_{\nu }\frac{1}{\not{k}+{\not{k}}_{2}-m}\gamma _{5}\frac{1}{\not{k}+{\not{k}}%
_{3}-m}\gamma _{\mu }\frac{1}{\not{k}+{\not{k}}_{1}-m}\right\} .
\end{eqnarray}
Taking the traces and integrating both sides in $k$ we can identify the
relation among the amplitudes 
\begin{equation}
\left( k_{3}-k_{2}\right) _{\lambda }T_{\lambda \mu \nu }^{AVV}\left(
k_{1},k_{2},k_{3};m\right) =-2miT_{\mu \nu }^{PVV}\left(
k_{1},k_{2},k_{3};m\right) -T_{\nu \mu }^{AV}\left( k_{3},k_{1};m\right)
+T_{\mu \nu }^{AV}\left( k_{1},k_{2};m\right) .
\end{equation}
So the symmetry relations that relate the processes $A\rightarrow VV$ and ${%
P\rightarrow VV}$ become 
\begin{equation}
q_{\lambda }T_{\lambda \mu \nu }^{A\rightarrow VV}=-2miT_{\mu \nu
}^{P\rightarrow VV}+T_{\mu \nu }^{AV}\left( k_{1},k_{2};m\right) -T_{\nu \mu
}^{AV}\left( k_{3},k_{1};m\right) +T_{\nu \mu }^{AV}\left(
l_{1},l_{2};m\right) -T_{\mu \nu }^{AV}\left( l_{3},l_{1};m\right) ,
\end{equation}
where the crossed channel was included. The above relation can easily be
extracted from the current algebra procedure. It must be noted that the Ward
identity can be satisfied, if and only if all four integrals on the right
side vanish identically. We identified such integrals with the previously
studied two-point function $T_{\mu \nu }^{AV}$. Substituting the results
obtained we get the expression 
\begin{eqnarray}
q_{\lambda }T_{\lambda \mu \nu }^{A\rightarrow VV} &=&-2miT_{\mu \nu
}^{P\rightarrow VV}  \nonumber \\
&&+2\varepsilon _{\mu \nu \alpha \beta }\left[ (k_{1}-k_{3})_{\beta
}(k_{1}+k_{3})_{\xi }+(k_{2}-k_{1})_{\beta }(k_{1}+k_{2})_{\xi }\right]
\triangle _{\xi \alpha }  \nonumber \\
&&-2\varepsilon _{\mu \nu \alpha \beta }\left[ (l_{1}-l_{3})_{\beta
}(l_{1}+l_{3})_{\xi }+(l_{2}-l_{1})_{\beta }(l_{1}+l_{2})_{\xi }\right]
\triangle _{\xi \alpha }.
\end{eqnarray}
Following the same strategy, many other relations can be constructed. In all
cases we simply substitute the results obtained for the corresponding
two-point functions after the symmetry relation for the three-point
functions has been constructed. To be as brief as possible we simply quote
the remaining results; 
\begin{eqnarray}
\bullet p_{\mu }T_{\lambda \mu \nu }^{A\rightarrow VV} &=&2\varepsilon
_{\lambda \nu \alpha \beta }\left[ (k_{2}-k_{1})_{\beta }(k_{1}+k_{2})_{\xi
}+(k_{3}-k_{2})_{\beta }(k_{2}+k_{3})_{\xi }\right] \left[ \triangle _{\xi
\alpha }\right]   \nonumber \\
&&+2\varepsilon _{\lambda \nu \alpha \beta }\left[ (l_{3}-l_{1})_{\beta
}(l_{1}+l_{3})_{\xi }+(l_{2}-l_{3})_{\beta }(l_{2}+l_{3})_{\xi }\right] %
\left[ \triangle _{\xi \alpha }\right]  \\
\bullet p_{\nu }^{\prime }T_{\lambda \mu \nu }^{A\rightarrow VV}
&=&2\varepsilon _{\lambda \mu \alpha \beta }\left[ (k_{3}-k_{1})_{\beta
}(k_{1}+k_{3})_{\xi }+(k_{2}-k_{3})_{\beta }(k_{2}+k_{3})_{\xi }\right] %
\left[ \triangle _{\xi \alpha }\right]   \nonumber \\
&&+2\varepsilon _{\lambda \mu \alpha \beta }\left[ (l_{2}-l_{1})_{\beta
}(l_{1}+l_{2})_{\xi }+(l_{3}-l_{2})_{\beta }(l_{2}+l_{3})_{\xi }\right] %
\left[ \triangle _{\xi \alpha }\right]  \\
\bullet q_{\lambda }T_{\lambda }^{V\rightarrow SS} &=&2(k_{2\alpha
}k_{2\beta }-k_{3\alpha }k_{3\beta })\left[ \triangle _{\alpha \beta }\right]
+2(l_{2\alpha }l_{2\beta }-l_{3\alpha }l_{3\beta })\left[ \triangle _{\alpha
\beta }\right]  \\
\bullet q_{\lambda }T_{\lambda }^{V\rightarrow PP} &=&2(k_{3\alpha
}k_{3\beta }-k_{2\alpha }k_{2\beta })\left[ \triangle _{\alpha \beta }\right]
+2(l_{3\alpha }l_{3\beta }-l_{2\alpha }l_{2\beta })\left[ \triangle _{\alpha
\beta }\right]  \\
\bullet q_{\lambda }T_{\lambda }^{A\rightarrow SP} &=&-2mi[T^{P\rightarrow
SP}]-2i(k_{3\alpha }k_{3\beta }-k_{2\alpha }k_{2\beta })\left[ \triangle
_{\alpha \beta }\right] -2i(l_{3\alpha }l_{3\beta }-l_{2\alpha }l_{2\beta })%
\left[ \triangle _{\alpha \beta }\right]  \\
\bullet p_{\mu }T_{\mu \nu }^{S\rightarrow VV} &=&4m(k_{3}-k_{1})_{\xi }
\left[ \triangle _{\xi \nu }\right] +4m(l_{1}-l_{2})_{\xi }\left[ \triangle
_{\xi \nu }\right] =8mp_{\xi }\left[ \triangle _{\xi \nu }\right]  \\
\bullet p_{\nu }^{\prime }T_{\mu \nu }^{S\rightarrow VV}
&=&4m(k_{1}-k_{2})_{\xi }\left[ \triangle _{\xi \mu }\right]
+4m(l_{3}-l_{1})_{\xi }\left[ \triangle _{\xi \mu }\right] =8mp_{\xi
}^{\prime }\left[ \triangle _{\xi \mu }\right]  \\
\bullet p_{\mu }T_{\mu \nu }^{S\rightarrow AA} &=&2mi[T_{\nu }^{S\rightarrow
PA}]-4m(k_{3}+k_{2})_{\xi }\left[ \triangle _{\xi \nu }\right]
+4m(l_{3}+l_{2})_{\xi }\left[ \triangle _{\xi \nu }\right]  \\
\bullet p_{\nu }^{\prime }T_{\mu \nu }^{S\rightarrow AA} &=&2mi[T_{\mu
}^{S\rightarrow AP}]+4m(k_{3}+k_{2})_{\xi }\left[ \triangle _{\xi \mu }%
\right] -4m(l_{3}+l_{2})_{\xi }\left[ \triangle _{\xi \mu }\right]  \\
\bullet q_{\lambda }T_{\lambda \mu }^{V\rightarrow AP} &=&0 \\
\bullet p_{\mu }T_{\lambda \mu }^{V\rightarrow AP} &=&2mi[T_{\lambda
}^{V\rightarrow PP}]-4m(k_{2}+k_{3})_{\alpha }\left[ \triangle _{\alpha
\lambda }\right] -4m(l_{2}+l_{3})_{\alpha }\left[ \triangle _{\alpha \lambda
}\right]  \\
\bullet q_{\lambda }T_{\lambda \mu \nu }^{A\rightarrow AA} &=&-2miT_{\mu \nu
}^{P\rightarrow AA}  \nonumber \\
&&+2\varepsilon _{\mu \nu \alpha \beta }\left[ (k_{3}-k_{1})_{\beta
}(k_{1}+k_{3})_{\xi }+(k_{1}-k_{2})_{\beta }(k_{1}+k_{2})_{\xi }\right] %
\left[ \triangle _{\xi \alpha }\right]   \nonumber \\
&&+2\varepsilon _{\mu \nu \alpha \beta }\left[ (l_{3}-l_{1})_{\beta
}(l_{1}+l_{3})_{\xi }+(l_{1}-l_{2})_{\beta }(l_{1}+l_{2})_{\xi }\right] %
\left[ \triangle _{\xi \alpha }\right]  \\
\bullet p_{\mu }T_{\lambda \mu \nu }^{A\rightarrow AA} &=&2miT_{\lambda \nu
}^{A\rightarrow PA}  \nonumber \\
&&-2\varepsilon _{\lambda \nu \alpha \beta }\left[ (k_{1}-k_{2})_{\beta
}(k_{1}+k_{2})_{\xi }+(k_{3}-k_{2})_{\beta }(k_{2}+k_{3})_{\xi }\right] %
\left[ \triangle _{\xi \alpha }\right]   \nonumber \\
&&+2\varepsilon _{\lambda \nu \alpha \beta }\left[ (l_{3}-l_{2})_{\beta
}(l_{1}+l_{3})_{\xi }+(l_{1}-l_{3})_{\beta }(l_{1}+l_{3})_{\xi }\right] %
\left[ \triangle _{\xi \alpha }\right]  \\
\bullet p_{\nu }^{\prime }T_{\lambda \mu \nu }^{A\rightarrow AA}
&=&2miT_{\lambda \mu }^{A\rightarrow AP}  \nonumber \\
&&+2\varepsilon _{\lambda \mu \alpha \beta }\left[ (k_{1}-k_{3})_{\beta
}(k_{1}+k_{3})_{\xi }+(k_{3}-k_{2})_{\beta }(k_{2}+k_{3})_{\xi }\right] %
\left[ \triangle _{\xi \alpha }\right]   \nonumber \\
&&+2\varepsilon _{\lambda \mu \alpha \beta }\left[ (l_{3}-l_{2})_{\beta
}(l_{2}+l_{3})_{\xi }+(l_{1}-l_{2})_{\beta }(l_{1}+l_{2})_{\xi }\right] %
\left[ \triangle _{\xi \alpha }\right] 
\end{eqnarray}

\begin{eqnarray}
\bullet p_{\mu }T_{\lambda \mu \nu }^{V\rightarrow AA} &=&2miT_{\lambda \nu
}^{V\rightarrow PA}  \nonumber \\
&&+\left[ (l_{1}+l_{3})_{\alpha }(l_{1}+l_{3})_{\beta
}-(l_{2}+l_{3})_{\alpha }(l_{2}+l_{3})_{\beta }\right] \left[ \Box _{\alpha
\beta \lambda \nu }-g_{\lambda \beta }\triangle _{\nu \alpha }-g_{\lambda
\alpha }\triangle _{\nu \beta }-3g_{\lambda \nu }\triangle _{\alpha \beta }%
\right]  \nonumber \\
&&-\left[ (k_{1}+k_{2})_{\alpha }(k_{1}+k_{2})_{\beta
}+(k_{2}+k_{3})_{\alpha }(k_{2}+k_{3})_{\beta }\right] \left[ \Box _{\alpha
\beta \lambda \nu }-g_{\lambda \beta }\triangle _{\nu \alpha }-g_{\lambda
\alpha }\triangle _{\nu \beta }-3g_{\lambda \nu }\triangle _{\alpha \beta }%
\right]  \nonumber \\
&&+\frac{2}{3}\left[ k_{2\alpha }(k_{1}+k_{3})_{\beta }-k_{2\beta
}(k_{1}+k_{2})_{\alpha }+l_{3\alpha }(l_{1}-l_{2})_{\beta }-l_{3\beta
}(l_{1}-l_{2})_{\alpha }\right] \left[ \Box _{\alpha \beta \lambda \nu
}+g_{\nu \alpha }\triangle _{\lambda \beta }+g_{\lambda \alpha }\triangle
_{\nu \beta }\right] .
\end{eqnarray}
And finally, 
\begin{eqnarray}
\bullet q_{\lambda }T_{\lambda \mu \nu }^{V\rightarrow VV} &=&\left[
(k_{1}+k_{2})_{\alpha }(k_{1}+k_{2})_{\beta }-(k_{1}+k_{3})_{\alpha
}(k_{1}+k_{3})_{\beta }\right] \left[ \Box _{\alpha \beta \mu \nu }-g_{\mu
\alpha }\triangle _{\nu \beta }-g_{\nu \alpha }\triangle _{\mu \beta
}-3g_{\mu \nu }\triangle _{\alpha \beta }\right]  \nonumber \\
&&+\left[ (l_{1}+l_{2})_{\alpha }(l_{1}+l_{2})_{\beta
}-(l_{1}+l_{3})_{\alpha }(l_{1}+l_{3})_{\beta }\right] \left[ \Box _{\alpha
\beta \mu \nu }-g_{\mu \alpha }\triangle _{\nu \beta }-g_{\nu \alpha
}\triangle _{\mu \beta }-3g_{\mu \nu }\triangle _{\alpha \beta }\right] 
\nonumber \\
&&+\frac{2}{3}\left[ k_{1\beta }(k_{3}-k_{2})_{\alpha }-k_{1\alpha
}(k_{3}-k_{2})_{\beta }+l_{1\beta }(l_{3}-l_{2})_{\alpha }-l_{1\alpha
}(l_{3}-l_{2})_{\beta }\right] \left[ \Box _{\alpha \beta \mu \nu }+g_{\mu
\alpha }\triangle _{\nu \beta }+g_{\nu \alpha }\triangle _{\mu \beta }\right]
.
\end{eqnarray}

Let us consider the values for our divergent basic objects from the point of
view of very representative techniques to handle divergences, in order to
then perform the analysis. We refer to the remaining objects in the
calculated expressions for the amplitudes and the symmetry relations
considered, since these values assumed a crucial role in the Ward identity
preservation as shown in eqs.(48)-(64).

\section{Basic Divergent Objects versus Regularization}

In the preceding sections, taking specific examples of sets of amplitudes,
we considered the aspects of ambiguities and symmetry relations. All the
conditions were put in terms of three differences among divergent integrals
of the same degree of divergence. From the results obtained by our
calculational strategy we can get the corresponding ones furnished by
traditional regularization methods or equivalent philosophies. All we need
is to specify the values attributed to the objects (4)-(8) by the  method
adopted. In this section we want to analyze how representative treatments of
divergences specify the values of these objects.

\subsection{Dimensional regularization}

For the evaluation of the momentum integrals in the DR technique \cite
{LEIBRANDT} we take as a starting point the validity of the expression 
\begin{equation}
I\left( 2\omega ,\alpha ,q\right) =\int \frac{d^{2\omega }k}{\left( 2\pi
\right) ^{2\omega }}\frac{1}{\left[ k^{2}+2q\cdot k-H^{2}\right] ^{\alpha }}%
=\left( \frac{i}{\left( 4\pi \right) ^{\omega }}\right) \frac{\Gamma \left(
\alpha -\omega \right) }{\Gamma \left( \alpha \right) \left(
-q^{2}-H^{2}\right) ^{\alpha -\omega }}.
\end{equation}
In situations where the divergent integrals are $(\alpha \leq \omega )$ we
admit that the integral $I\left( 2\omega ,\alpha ,q\right) $ is an analytic
function of the variable $\omega $ which is continuous and complex. On the
right hand side the Gamma function is changed from the Euler function to its
analytic continuation (in the region $\alpha \leq \omega $): the Weierstrass
function. The divergences will emerge as poles at specific values for $%
\omega $. An important aspect of our discussion is that once the result (65)
is clearly established we can use it to produce relations among integrals
without being concerned about the divergences. The specific relations of
interest here can be produced by the adequate differentiation of both sides
of eq.(65) relative to the $q$ momentum, and then, taking $Q$ as zero.
Following this procedure, we can find 
\begin{eqnarray}
\bullet \int \frac{d^{2\omega }k}{\left( 2\pi \right) ^{2\omega }}\frac{1}{%
\left( k^{2}-m^{2}\right) ^{\alpha }} &=&\left( \frac{i}{\left( 4\pi \right)
^{\omega }}\right) \frac{\Gamma \left( \alpha -\omega \right) }{\Gamma
\left( \alpha \right) \left( -m^{2}\right) ^{\alpha -\omega }}, \\
\bullet \int \frac{d^{2\omega }k}{\left( 2\pi \right) ^{2\omega }}\frac{%
k_{\mu }k_{\nu }}{\left( k^{2}-m^{2}\right) ^{\alpha }} &=&\left( \frac{i}{%
\left( 4\pi \right) ^{\omega }}\right) \frac{1}{2}g_{\mu \nu }\frac{\Gamma
\left( \alpha -\omega -1\right) }{\Gamma \left( \alpha \right) \left(
-m^{2}\right) ^{\alpha -\omega -1}}, \\
\bullet \int \frac{d^{2\omega }k}{\left( 2\pi \right) ^{2\omega }}\frac{%
k_{\mu }k_{\nu }k_{\xi }k_{\beta }}{\left( k^{2}-m^{2}\right) ^{\alpha }}
&=&\left( \frac{i}{\left( 4\pi \right) ^{\omega }}\right) \frac{1}{4}\left[
g_{\mu \nu }g_{\beta \xi }+g_{\mu \beta }g_{\nu \xi }+g_{\mu \xi }g_{\nu
\beta }\right] \frac{\Gamma \left( \alpha -\omega -2\right) }{\Gamma \left(
\alpha \right) \left( -m^{2}\right) ^{\alpha -\omega -2}}.
\end{eqnarray}
The comparison with the results can be used to identify the relations 
\begin{eqnarray}
\bullet \int \frac{d^{2\omega }k}{\left( 2\pi \right) ^{2\omega }}\frac{%
2k_{\mu }k_{\nu }}{\left( k^{2}-m^{2}\right) ^{2}} &=&\int \frac{d^{2\omega
}k}{\left( 2\pi \right) ^{2\omega }}\frac{g_{\mu \nu }}{\left(
k^{2}-m^{2}\right) }, \\
\bullet \int \frac{d^{2\omega }k}{\left( 2\pi \right) ^{2\omega }}\frac{%
4k_{\mu }k_{\nu }}{\left( k^{2}-m^{2}\right) ^{3}} &=&\int \frac{d^{2\omega
}k}{\left( 2\pi \right) ^{2\omega }}\frac{g_{\mu \nu }}{\left(
k^{2}-m^{2}\right) ^{2}}, \\
\bullet \int \frac{d^{2\omega }k}{\left( 2\pi \right) ^{4}}\frac{24k_{\nu
}k_{\mu }k_{\beta }k_{\xi }}{\left( k^{2}-m^{2}\right) ^{4}} &=&g_{\mu \nu
}\int \frac{d^{2\omega }k}{\left( 2\pi \right) ^{4}}\frac{4k_{\beta }k_{\xi }%
}{\left( k^{2}-m^{2}\right) ^{3}}+g_{\mu \beta }\int \frac{d^{2\omega }k}{%
\left( 2\pi \right) ^{4}}\frac{4k_{\xi }k_{\nu }}{\left( k^{2}-m^{2}\right)
^{3}}+g_{\nu \beta }\int \frac{d^{2\omega }k}{\left( 2\pi \right) ^{4}}\frac{%
4k_{\xi }k_{\mu }}{\left( k^{2}-m^{2}\right) ^{3}}.
\end{eqnarray}
So the consistency conditions that emerged in our analysis are automatically
satisfied in the DR technique. It is possible to say that the consistency of
the DR scheme resides precisely in this fact. The results produced by the
technique are automatically free from ambiguities and are symmetry
preserving. It is allowed to perform shifts on the integrating momentum, but
it is not a compulsory operation. The properties (69)-(71) eliminate all the
possible dependence on the choices of internal line momenta. Because of this
conclusion it is easy to see that in all problems where the DR technique can
be applied if we take $\Box _{\alpha \beta \mu \nu }=\nabla _{\mu \nu
}=\triangle _{\mu \nu }=0$ and write the objects $I_{log}\left( m^{2}\right) 
$ and $I_{quad}\left( m^{2}\right) $ [taking in their coefficients the
appropriate values for the traces of the $\gamma $ matrices $tr\left( \gamma
_{\mu }\gamma _{\nu }\right) =2^{\omega }g_{\mu \nu }$, and so on] according
to the expressions directly dictated by eq.(65), a perfect map can be
obtained.

\subsection{The Pauli-Villars covariant regularization}

To evaluate any divergent integral from the point of view of the
Pauli-Villars (PV) prescription \cite{PV}, we initially take the
substitution 
\begin{equation}
I(m)\longrightarrow \sum_{i=0}a_{i}I(\Lambda _{i}),
\end{equation}
where $a_{0}=1$ and $\Lambda _{0}=m$. All the other $a_{i}$ and $\Lambda _{i}
$ parameters need to be chosen in such a way as to construct a superposition
leading to the desirable results, guided, for example, by the maintenance of
the Ward identity. In terms of this recipe let us consider the values of the
three relevant differences. First, 
\begin{eqnarray}
&&\!\!\!\!\!\!\!\!\!\!\left\{ \int_{\Lambda }\frac{d^{4}k}{\left( 2\pi
\right) ^{4}}\frac{4k_{\mu }k_{\nu }}{\left( k^{2}-m^{2}\right) ^{3}}%
-\int_{\Lambda }\frac{d^{4}k}{\left( 2\pi \right) ^{4}}\frac{g_{\mu \nu }}{%
\left( k^{2}-m^{2}\right) ^{2}}\right\} =\sum_{i=0}a_{i}\left\{
\int_{\Lambda }\frac{d^{4}k}{\left( 2\pi \right) ^{4}}\frac{4k_{\mu }k_{\nu }%
}{\left( k^{2}-\Lambda _{i}^{2}\right) ^{3}}-\int_{\Lambda }\frac{d^{4}k}{%
\left( 2\pi \right) ^{4}}\frac{g_{\mu \nu }}{\left( k^{2}-\Lambda
_{i}^{2}\right) ^{2}}\right\}   \nonumber \\
&&\;\;\;\;\;\;\;\;\;\;\;\;\;\;\;\;\;\;\;\;\;\;\;\;\;\;\;\;\;\;\;\;\;\;\;\;=%
\sum_{i=0}a_{i}\left\{ \int_{\Lambda }\frac{d^{4}k}{\left( 2\pi \right) ^{4}}%
\frac{\Lambda _{i}^{2}}{\left( k^{2}-\Lambda _{i}^{2}\right) ^{3}}\right\}
\;=\sum_{i=0}a_{i}\left\{ \left( \frac{i}{(4\pi )^{2}}\right) \left( -\frac{1%
}{2}\right) \right\} .
\end{eqnarray}
To satisfy the above consistency condition all we need is to choose a set of 
$a_{i}^{\prime }s$ so that $\sum_{i=0}a_{i}=0$. The next condition, the one
relative to $\Box _{\alpha \beta \mu \nu }$, leads to the result 
\begin{equation}
\Box _{\alpha \beta \mu \nu }=\left[ g_{\mu \nu }g_{\alpha \beta }+g_{\alpha
\mu }g_{\nu \beta }+g_{\alpha \nu }g_{\beta \mu }\right] \sum_{i=0}a_{i}%
\left\{ \left( \frac{i}{(4\pi )^{2}}\right) \left( -\frac{5}{6}\right)
\right\} ,
\end{equation}
which is then simultaneously satisfied with the same choice of coefficients
as the one previously considered. The relation involving quadratic
divergences ($\nabla _{\mu \nu }$) can be evaluated in the same way: 
\begin{eqnarray}
\nabla _{\mu \nu } &=&\sum_{i=0}a_{i}\left\{ \int_{\Lambda }\frac{d^{4}k}{%
\left( 2\pi \right) ^{4}}\frac{2k_{\mu }k_{\nu }}{\left( k^{2}-\Lambda
_{i}^{2}\right) ^{2}}-\int_{\Lambda }\frac{d^{4}k}{\left( 2\pi \right) ^{4}}%
\frac{g_{\mu \nu }}{\left( k^{2}-\Lambda _{i}^{2}\right) }\right\}  
\nonumber \\
&&\!\!\!\!\!\!\!\!\!\!\!\!\!\!\!\!\!\!=-\frac{g_{\mu \nu }}{2}%
\sum_{i=0}a_{i}\int_{\Lambda }\frac{d^{4}k}{\left( 2\pi \right) ^{4}}\frac{1%
}{\left( k^{2}-\Lambda _{i}^{2}\right) }+\frac{g_{\mu \nu }}{2}%
\sum_{i=0}a_{i}\Lambda _{i}^{2}\int_{\Lambda }\frac{d^{4}k}{\left( 2\pi
\right) ^{4}}\frac{1}{\left( k^{2}-\Lambda _{i}^{2}\right) ^{2}}.
\end{eqnarray}
So if we choose, in addition to the choice $\sum_{i}a_{i}=0$, the values for 
$a_{i}$ and $\Lambda _{i}^{2}$ so that $\sum_{i=0}a_{i}\Lambda _{i}^{2}=0$,
we get $\nabla _{\mu \nu }=0$. There are no new facts in the conditions
derived above. Actually, they are the same ones used in the QED treatment,
guided by the maintenance of gauge invariance in the vacuum polarization
tensor, in the vanishing of the tadpole diagram of the electron self-energy,
and so on. To complete this subsection we need to show how the Pauli-Villars
results can be extracted from ours. The first step is obviously to take $%
\Box _{\alpha \beta \mu \nu }=\nabla _{\mu \nu }=\triangle _{\mu \nu }=0$,
and then evaluate the remaining divergent objects according to the PV
prescription with the conditions $\sum_{i}a_{i}=0$, and $\sum_{i=0}a_{i}%
\Lambda _{i}^{2}=0$ dictated by the above investigations.

Explicitly, for the quadratic divergent object we have 
\begin{eqnarray}
\int_{\Lambda }\frac{d^{4}k}{\left( 2\pi \right) ^{4}}\frac{1}{\left(
k^{2}-m^{2}\right) } &=&\sum_{i=0}a_{i}\int \frac{d^{4}k}{\left( 2\pi
\right) ^{4}}\frac{1}{\left( k^{2}-\Lambda _{i}^{2}\right) }=\int_{\Lambda }%
\frac{d^{4}k}{\left( 2\pi \right) ^{4}}\left\{ \frac{a_{0}}{k^{2}-\Lambda
_{0}^{2}}+\frac{a_{1}}{k^{2}-\Lambda _{1}^{2}}+\frac{a_{2}}{k^{2}-\Lambda
_{2}^{2}}\right\}   \nonumber \\
&=&\left. \int \frac{d^{4}k}{\left( 2\pi \right) ^{4}}\right\{
k^{4}(a_{0}+a_{1}+a_{2})-k^{2}[a_{0}(\Lambda _{1}^{2}+\Lambda
_{2}^{2})+a_{1}(\Lambda _{0}^{2}+\Lambda _{2}^{2})+a_{2}(\Lambda
_{0}^{2}+\Lambda _{1}^{2})]  \nonumber \\
&&\left. +a_{0}\Lambda _{1}^{2}\Lambda _{2}^{2}+a_{1}\Lambda _{0}^{2}\Lambda
_{2}^{2}+a_{2}\Lambda _{1}^{2}\Lambda _{0}^{2}\right\} \left\{ \frac{1}{%
(k^{2}-\Lambda _{0}^{2})(k^{2}-\Lambda _{1}^{2})(k^{2}-\Lambda _{2}^{2})}%
\right\} .
\end{eqnarray}
The condition $\sum_{i}a_{i}=0$ appears in the $k^{4}$ coefficient, which is
the highest power of the integrated momentum. The condition $%
\sum_{i=0}a_{i}\Lambda _{i}^{2}=0$ is present in the $k^{2}$ coefficient.
Even though the choices we make were dictated by our consistency conditions,
it is interesting to note that they are obvious ingredients in order to
guarantee the regularizability of the corresponding divergent integral. With
this recipe we need to add powers of $k^{2}$ only in the denominator to
modify the behavior in the region of high values of the integrating
momentum. Solving the equations we get the expression 
\begin{equation}
\int_{\Lambda }\frac{d^{4}k}{\left( 2\pi \right) ^{4}}\frac{1}{\left(
k^{2}-m^{2}\right) }\longrightarrow \int \frac{d^{4}k}{\left( 2\pi \right)
^{4}}\frac{1}{\left( k^{2}-m^{2}\right) }\frac{(m^{2}-\Lambda _{1}^{2})}{%
(k_{2}^{2}-\Lambda _{1}^{2})}\frac{(m^{2}-\Lambda _{2}^{2})}{%
(k_{2}^{2}-\Lambda _{2}^{2})},
\end{equation}
and because of the finite character of the modified integral the solution is
immediate.

\subsection{Surface term evaluation}

In the literature about divergences in QFT there is another procedure which
plays an important role: the explicit evaluation of surface terms. They are
involved when shifts in the integrated momentum are performed to relate one
specific choice for the internal line momenta with another one. This is
frequently used in the verification of the Ward identities involving
situations where DR cannot be used. The most remarkable one is the
justification of the violations of symmetry relations in the $AVV$ triangle
amplitude \cite{JACKIW1}\cite{JACKIW2}. Because in our calculations the
choices for the routing of internal line momenta were adopted as the most
general ones, and no shifts were taken in the performing steps, it is
possible to map our results onto those corresponding to this specific
procedure. All we need is to relate the consistency conditions to the
surface terms. This is not a difficult job if we note a small number of
relations such as 
\begin{equation}
\frac{\partial }{\partial k_{\nu }}\left( \frac{k_{\mu }}{\left(
k^{2}-m^{2}\right) ^{\alpha }}\right) =\frac{g_{\mu \nu }}{\left(
k^{2}-m^{2}\right) ^{\alpha }}-2\alpha \frac{k_{\mu }k_{\nu }}{\left(
k^{2}-m^{2}\right) ^{\alpha +1}}.
\end{equation}
So, if we integrate both sides in the momentum $k$, we have 
\begin{equation}
\int \frac{d^{4}k}{\left( 2\pi \right) ^{4}}\frac{\partial }{\partial k_{\nu
}}\left( \frac{k_{\mu }}{\left( k^{2}-m^{2}\right) ^{\alpha }}\right) =\int 
\frac{d^{4}k}{\left( 2\pi \right) ^{4}}\frac{g_{\mu \nu }}{\left(
k^{2}-m^{2}\right) ^{\alpha }}-2\alpha \int \frac{d^{4}k}{\left( 2\pi
\right) ^{4}}\frac{k_{\mu }k_{\nu }}{\left( k^{2}-m^{2}\right) ^{\alpha +1}}.
\end{equation}
Now, taking $\alpha =1$ and $\alpha =2,$ respectively, we obtain 
\begin{eqnarray}
\bullet \int \frac{d^{4}k}{\left( 2\pi \right) ^{4}}\frac{\partial }{%
\partial k_{\nu }}\left( \frac{k_{\mu }}{\left( k^{2}-m^{2}\right) }\right) 
&=&-\nabla _{\mu \nu } \\
\bullet \int \frac{d^{4}k}{\left( 2\pi \right) ^{4}}\frac{\partial }{%
\partial k_{\nu }}\left( \frac{k_{\mu }}{\left( k^{2}-m^{2}\right) ^{2}}%
\right)  &=&-\triangle _{\mu \nu },
\end{eqnarray}
and following the same procedure we can state that 
\begin{equation}
\int \frac{d^{4}k}{\left( 2\pi \right) ^{4}}\frac{\partial }{\partial k_{\nu
}}\left( \frac{k_{\mu }k_{\alpha }k_{\beta }}{\left( k^{2}-m^{2}\right) ^{3}}%
\right) =-\Box _{\alpha \beta \mu \nu }.
\end{equation}
On the left hand side of the last three equations we have an integral of a
total derivative which is a surface term in consequence of the Gauss
theorem. When the integrand is convergent or logarithmically divergent in
the region of high values for the momentum, the surface terms vanish
identically; otherwise they do not. We can evaluate the values for the
surface terms related to the objects $\Box _{\alpha \beta \mu \nu },\nabla
_{\mu \nu },$ and $\triangle _{\mu \nu }$. For example $\triangle _{\mu \nu
}=-\frac{ig_{\mu \nu }}{32\pi ^{2}}$ and $\Box _{\alpha \beta \mu \nu }=%
\left[ g_{\mu \nu }g_{\alpha \beta }+g_{\alpha \mu }g_{\nu \beta }+g_{\alpha
\nu }g_{\beta \mu }\right] \frac{i}{(4\pi )^{2}}\left( -\frac{5}{6}\right) $%
. These values can be found in many papers or textbooks about triangle
anomalies. It is important to note that if this interpretation is adopted
the values for $\triangle _{\mu \nu }$ and $\Box _{\alpha \beta \mu \nu }$
are completely well defined and finite. The main aspect is that in this
specific point of view what we denominate the consistency condition is not
satisfied. Consequently, the amplitudes we calculated in sections III and IV
are not free from ambiguities and the Ward identities are not automatically
satisfied. This way of looking at divergences in perturbative calculations
of QFT is clearly not compatible with that of DR and the PV technique.

\section{Final Remarks and Conclusions}

The divergent content of all the considered amplitudes and their respective
Ward identities can be put in terms of only five basic objects, eqs.(4)-(8).
The strategy adopted to handle the divergent Feynman integrals preserves all
arbitrariness and allows us to emphasize exactly where choices have to be
made in the perturbative calculations, in order to obtain a definite value
for the physical amplitudes. The results obtained within the adopted
strategy can be mapped onto the corresponding results of any traditional
regularization technique or equivalent philosophy. In this framework it
becomes possible to state clearly the relevant issues concerning
regularizations, ambiguities, and symmetry relations.

Regularizations can be classified into two basic classes. In the first class
we can put the regularizations obeying the consistency conditions 
\begin{equation}
\Box _{\alpha \beta \mu \nu }^{reg}=\nabla _{\mu \nu }^{reg}=\triangle _{\mu
\nu }^{reg}=0.
\end{equation}
In this class we find the popular and convenient DR technique. A second
class of regularizations can be characterized by 
\begin{equation}
\left( \Box _{\alpha \beta \mu \nu }^{reg};\nabla _{\mu \nu
}^{reg};\triangle _{\mu \nu }^{reg}\right) \neq 0.
\end{equation}
In this class we find the surface term evaluation analysis that is commonly
found in the literature devoted to the triangle anomalies (where DR cannot
be straightforwardly applied).

The two classes defined above lead to drastically different descriptions of
perturbative amplitudes in two respects.

$i)$ Ambiguities: since the potentially ambiguous terms of all amplitudes
are always multiplied by one of the objects (4)-(6), the first class of
regularizations will completely eliminate the ambiguities. For the second
class, physical amplitudes will become ambiguous so that we will have to
make choices for the nonphysical quantities. The predictive power of QFT is
clearly affected; predictions cannot be made in a definite way, fundamental
symmetries like the space-time homogeneity may be broken, and so on.

$ii)$ Symmetry relations: We noticed that the potential symmetry violating
terms are always multiplied by one of the objects (4)-(6), like the
ambiguous terms. However, note that all the ambiguous terms are symmetry
violating ones, but the symmetry violating terms are not always ambiguous.
The last sentence means that the second class of regularizations cannot lead
us to consistent results just because there are cases where no choices can
avoid the violations.

At this point, if one wants to look at all QFT perturbative problems in the
same way, one must consider the question of triangle anomalies. It seems
that there are two situations which should be taken into account.

$i)$ In the first class of regularizations, apparently all amplitudes become
automatically unambiguous and symmetry preserving. In particular, the right
hand sides of eqs.(48)-(50) and (60)-(62) seem to imply that no violations
of axial symmetry relations can be obtained. How, then, can we recover the
triangle anomalies in this class of regularizations? We return to this
crucial question in a moment.

$ii)$ In the second class of regularizations, interpreting the objects
(4)-(6) as surface terms, we can get violations of symmetry relations in the 
$AVV$ and $AAA$ triangle amplitudes. This is precisely the traditional
procedure used to justify the anomalies. It is easy to show from our results
that such a description is immediately recovered. For this purpose it is
enough to get the value for the $\triangle _{\mu \nu }$ object dictated by
its surface's term interpretation; $\triangle _{\mu \nu }=-\frac{ig_{\mu \nu
}}{32\pi ^{2}}.$ Taking specifically the $AVV$ case, we assume, as usual, $%
k_{1},k_{2},$ and $k_{3}$, which are the arbitrary choices for the internal
line momenta, to be linear combinations of the physical momenta of the
external vectors, namely, $p$ and $p^{\prime }$, as follows 
\begin{equation}
\left\{ 
\begin{array}{l}
k_{1}=ap^{\prime }+bp \\ 
k_{2}=bp+(a-1)p^{\prime } \\ 
k_{3}=ap^{\prime }+(b+1)p.
\end{array}
\right. 
\end{equation}
Notice that $k_{1}-k_{2}=p^{\prime },k_{3}-k_{1}=p,$ and $%
k_{3}-k_{2}=p^{\prime }+p=q$, where $q$ is the momentum of the axial vector.
The results corresponding to the contribution of the direct channel in
eqs.(48)-(50) become 
\begin{eqnarray}
\bullet q_{\lambda }T_{\lambda \mu \nu }^{AVV} &=&-2miT_{\mu \nu }^{PVV}+%
\frac{\left( a-b\right) }{8\pi ^{2}}i\varepsilon _{\mu \nu \alpha \beta
}p^{\prime \alpha }p^{\beta }, \\
\bullet p_{\mu }T_{\lambda \mu \nu }^{AVV} &=&\frac{\left( 1-a\right) }{8\pi
^{2}}i\varepsilon _{\lambda \nu \alpha \beta }p^{\prime \alpha }p^{\beta },
\\
\bullet p_{\nu }^{\prime }T_{\lambda \mu \nu }^{AVV} &=&-\frac{\left(
1+b\right) }{8\pi ^{2}}i\varepsilon _{\lambda \mu \alpha \beta }p^{\prime
\alpha }p^{\beta }.
\end{eqnarray}
Note that there are no choices for the parameters $a$ and $b$ such that the
violating terms are simultaneously eliminated. Closer contact with the usual
results can be obtained if the contribution of the crossed diagram is added,
assuming the same significance for the arbitrary internal momenta, i.e., $k$%
's equal to $l$'s in eqs.(48)-(50). Then we get 
\begin{eqnarray}
\bullet q_{\lambda }T_{\lambda \mu \nu }^{A\rightarrow VV} &=&-2miT_{\mu \nu
}^{P\rightarrow VV}+\frac{\left( a-b\right) }{4\pi ^{2}}i\varepsilon _{\mu
\nu \alpha \beta }p^{\prime \alpha }p^{\beta }, \\
\bullet p_{\mu }T_{\lambda \mu \nu }^{A\rightarrow VV} &=&\frac{\left(
b-a+2\right) }{8\pi ^{2}}i\varepsilon _{\lambda \nu \alpha \beta }p^{\prime
\alpha }p^{\beta }, \\
\bullet p_{\nu }^{\prime }T_{\lambda \mu \nu }^{A\rightarrow VV} &=&-\frac{%
\left( b-a+2\right) }{8\pi ^{2}}i\varepsilon _{\lambda \mu \alpha \beta
}p^{\prime \alpha }p^{\beta }.
\end{eqnarray}
Putting $a=1$, the above results can be recognized as the usual ones \cite
{JACKIW1} -\cite{CHENG-LI}. So, adopting this point of view, we have a
justification for the violation of at least one of the above symmetry
relations, because there is no value for the arbitrary parameter $b$ such
that all the Ward identities can be simultaneously satisfied. However, as we
have discussed, this procedure will lead us to many undesirable features in
other physical amplitudes; namely ambiguities and violations of nonanomalous
symmetry relations. In particular, this justification for the triangle
anomalies is founded in a nonzero value for the $AV$ two-point function, as
can be seen from eqs.(48)-(50) and (38)-(40). However, the $AV$ two-point
function must be identically zero for many reasons. Therefore the procedure
is in some sense not consistent.

In view of the above statements, one might conclude that it is not possible
to treat all the situations involving divergences in QFT from a unique
perspective, and that one has to adopt a case-by-case approach. This would
mean that, in the nonanomalous amplitudes, we must apply a regularization
belonging to the first class, like DR, eliminating all the ambiguities,
while in the treatment of anomalous amplitudes we must use the second class
of regularizations. This seems to be the present status of the problem in
the literature. We find this attitude unaceptable because it implies that
identical mathematical structures can receive different values in different
amplitudes. Even though the choice of regularization is arbitrary, to use
two types of such mathematical tools belonging to different classes in the
same problem could represent an excessive degree of arbitrariness.

Up to this point we have used the results produced within our general
framework only to show that there are aspects in the present status of the
treatment of divergences that could be questionable. However, we now want to
show that it is, in fact, possible to treat all the situations in a
consistent way by the same method. Taking into account what we have learned,
it is clear that only the first class of regularizations can provide us
consistent results, at least for the nonanomalous cases. We now need to
explain how triangle anomaly phenomena can be accounted for from this point
of view. The crucial issue is to get the expected results for the $AAA$ and $%
AVV$ triangle amplitudes, in spite of the ambiguity elimination as a
consequence of the imposition of the consistency conditions. The first step
is naturally to explain why the zero value for the $AV$ structure is not
incompatible with the anomaly phenomena. Actually the phenomenological
foundations of the triangle anomalies reside in the connection of the $AVV$
amplitude with the neutral electromagnetic pion decay. Through standard
methods of current algebra, the partial conservation of axial-vector current
hypothesis, and the LSZ formalism, four constraints are imposed on the $AVV$
amplitude. Three of them are the Ward identities related to eqs.(48)-(50)
and the fourth one is the low energy $\lim_{q_{\lambda }\rightarrow
0}q_{\lambda }T_{\lambda \mu \nu }^{A\rightarrow VV}=0.$ In the context of
the Sutherland-Veltman paradox, it was stated that it is not possible to
obtain in any calculational method an expression for the $AVV$ amplitude
such that all four constraints are simultaneously satisfied. This is a
property of the $AVV$ mathematical structure. Given these statements,
eqs.(48)-(50) reveal only that after the evaluation of the $AVV$ structure
and the contractions with the external momenta are performed, mathematical
structures identical to that of $AV$ should be identified. The low energy
limit and the three Ward identities must not be simultaneously satisfied
with or without the $AV$ corresponding terms. This means that even when the $%
AV$ (ambiguous) terms are eliminated through the consistency conditions it
is expected that the triangle anomaly phenomena should emerge in a natural
way in the problem. In order to verify the complete set of ingredients
present in the Sutherland-Veltman paradox, it becomes necessary to make an
explicit evaluation, within our calculational framework, of the $AVV$
mathematical structure. All the details involved will be presented elsewhere 
\cite{ORIMAR AVV-AAA}. The main results can be described as it follows. If
we adopt strictly the same strategy described in section II to manipulate
the Feynman integrals involved, an expression for the $AVV$ amplitude can be
obtained where all the eventual arbitrariness is still present. The
contractions with the external momenta can be written, for the contribution
of the direct channel, as 
\begin{eqnarray}
\bullet \left( k_{3}-k_{1}\right) _{\mu }T_{\lambda \mu \nu }^{AVV}
&=&\left( k_{3}-k_{1}\right) _{\mu }\Gamma _{\lambda \mu \nu }^{AVV} 
\nonumber \\
&&+\left( \frac{i}{8\pi ^{2}}\right) \varepsilon _{\nu \beta \lambda \xi
}\left( k_{3}-k_{1}\right) _{\xi }\left( k_{1}-k_{2}\right) _{\beta }, \\
\bullet \left( k_{1}-k_{2}\right) _{\nu }T_{\lambda \mu \nu }^{AVV}
&=&\left( k_{1}-k_{2}\right) _{\nu }\Gamma _{\lambda \mu \nu }^{AVV} 
\nonumber \\
&&-\left( \frac{i}{8\pi ^{2}}\right) \varepsilon _{\mu \beta \lambda \xi
}\left( k_{3}-k_{1}\right) _{\xi }\left( k_{1}-k_{2}\right) _{\beta }, \\
\bullet \left( k_{3}-k_{2}\right) _{\lambda }T_{\lambda \mu \nu }^{AVV}
&=&\left( k_{3}-k_{2}\right) _{\lambda }\Gamma _{\lambda \mu \nu }^{AVV} 
\nonumber \\
&&+\left( \frac{i}{4\pi ^{2}}\right) \varepsilon _{\mu \xi \nu \beta }\left(
k_{3}-k_{1}\right) _{\xi }\left( k_{1}-k_{2}\right) _{\beta }\left[
2m^{2}\xi _{00}\right] ,
\end{eqnarray}
where we have defined 
\begin{eqnarray}
\Gamma _{\lambda \mu \nu }^{AVV} &=&\varepsilon _{\mu \nu \beta \xi }\left[
\left( k_{2}-k_{1}\right) _{\beta }+\left( k_{3}-k_{1}\right) _{\beta }%
\right] \Delta _{\lambda \xi }  \nonumber \\
&&-\varepsilon _{\nu \lambda \beta \xi }\left( k_{2}-k_{3}\right) _{\beta
}\Delta _{\mu \xi }-\varepsilon _{\mu \lambda \beta \xi }\left(
k_{2}-k_{3}\right) _{\beta }\Delta _{\nu \xi }  \nonumber \\
&&+\varepsilon _{\mu \nu \lambda \alpha }\left[ \left( k_{1}+k_{2}\right)
_{\beta }+\left( k_{3}+k_{1}\right) _{\beta }\right] \Delta _{\alpha \beta },
\end{eqnarray}
and $\xi _{00}$ is the same finite integral that arises in the evaluation of
the $PVV$ three-point function: 
\begin{equation}
T_{\mu \nu }^{PVV}=\left( \frac{1}{4\pi ^{2}}\right) m\varepsilon _{\mu \nu
\alpha \beta }\left( k_{1}-k_{2}\right) _{\beta }\left( k_{3}-k_{1}\right)
_{\alpha }\left( \xi _{00}\right) .
\end{equation}
In addition, looking at eqs.(92)-(94), we can see that it is possible to
identify $AV$ two-point functions given by eq.(26). To make clear this
observation, note that 
\begin{equation}
\left( k_{3}-k_{2}\right) _{\lambda }\Gamma _{\lambda \mu \nu
}^{AVV}=2\varepsilon _{\mu \nu \alpha \beta }\left[ (k_{1}-k_{3})_{\beta
}(k_{1}+k_{3})_{\xi }+(k_{2}-k_{1})_{\beta }(k_{1}+k_{2})_{\xi }\right]
\triangle _{\xi \alpha }.
\end{equation}
With these identifications, eq.(94) can be put precisely in the form of
eq.(46) so that the corresponding relation between Green's functions is
preserved before any assumption about the involved arbitrariness. On the
other hand, given the results (92) and (93), it is easy to see that the
corresponding relations between Green's functions stated for the
contractions with the momenta $\left( k_{3}-k_{1}\right) _{\mu }$ and $%
\left( k_{1}-k_{2}\right) _{\nu }$, eqs.(49) and (50), respectively, are not
satisfied. Next, in order to be consistent in the perturbative calculations,
we must adopt $\triangle _{\xi \alpha }=0,$ eliminating then the $AV$
corresponding structures present on the right hand side of eqs.(92)-(94). We
get then 
\begin{eqnarray}
\bullet \left( k_{3}-k_{1}\right) _{\mu }T_{\lambda \mu \nu }^{AVV}
&=&\left( \frac{i}{8\pi ^{2}}\right) \varepsilon _{\nu \beta \lambda \xi
}\left( k_{3}-k_{1}\right) _{\xi }\left( k_{1}-k_{2}\right) _{\beta }, \\
\bullet \left( k_{1}-k_{2}\right) _{\nu }T_{\lambda \mu \nu }^{AVV}
&=&-\left( \frac{i}{8\pi ^{2}}\right) \varepsilon _{\mu \beta \lambda \xi
}\left( k_{3}-k_{1}\right) _{\xi }\left( k_{1}-k_{2}\right) _{\beta }, \\
\bullet \left( k_{3}-k_{2}\right) _{\lambda }T_{\lambda \mu \nu }^{AVV}
&=&-2mi\left\{ T_{\mu \nu }^{PVV}\right\} .
\end{eqnarray}
If we include the crossed diagram's contribution, we get 
\begin{eqnarray}
\bullet p_{\mu }T_{\lambda \mu \nu }^{A\rightarrow VV} &=&\left( \frac{i}{%
4\pi ^{2}}\right) \varepsilon _{\nu \beta \lambda \xi }p_{\xi }p_{\beta
}^{\prime }, \\
\bullet p_{\nu }^{\prime }T_{\lambda \mu \nu }^{A\rightarrow VV} &=&-\left( 
\frac{i}{4\pi ^{2}}\right) \varepsilon _{\mu \beta \lambda \xi }p_{\xi
}p_{\beta }^{\prime }, \\
\bullet q_{\lambda }T_{\lambda \mu \nu }^{A\rightarrow VV} &=&-2mi\left\{
T_{\mu \nu }^{P\rightarrow VV}\right\} .
\end{eqnarray}

The set of results obtained is in agreement with the statements of the
Sutherland-Veltman paradox. Only one of the four symmetry properties, the
axial Ward identity, was preserved. However, as is well known, the low
energy limit is related to the neutral electromagnetic pion decay. So, in
order for the $AVV$ amplitude to conform to the phenomenology, we have to
modify the calculated expression in an ad hoc way, by fixing the correct
behavior to the low energy limit by defining 
\begin{equation}
\left( T_{\lambda \mu \nu }^{A\rightarrow VV}(p,p^{\prime })\right)
_{phys}=T_{\lambda \mu \nu }^{A\rightarrow VV}(p,p^{\prime })-T_{\lambda \mu
\nu }^{A\rightarrow VV}\left( 0\right) ,
\end{equation}
where 
\begin{equation}
T_{\lambda \mu \nu }^{A\rightarrow VV}(0)=-\left( \frac{i}{4\pi ^{2}}\right)
\varepsilon _{\mu \nu \lambda \xi }\left[ p_{\xi }-p_{\xi }^{\prime }\right]
,
\end{equation}
is, as it should be, the required anomalous term. Consequently, we get for
the $AVV$ physical amplitude 
\begin{eqnarray}
\bullet p_{\nu }^{\prime }\left( T_{\lambda \mu \nu }^{A\rightarrow
VV}\right) _{phy} &=&0, \\
\bullet p_{\mu }\left( T_{\lambda \mu \nu }^{A\rightarrow VV}\right) _{phy}
&=&0, \\
\bullet q_{\lambda }\left( T_{\lambda \mu \nu }^{A\rightarrow VV}\right)
_{phy} &=&-2mi\left\{ T_{\mu \nu }^{P\rightarrow VV}\right\} -\left( \frac{i%
}{2\pi ^{2}}\right) \varepsilon _{\mu \nu \alpha \beta }\left[ p_{\alpha
}p_{\beta }^{\prime }\right] .
\end{eqnarray}

So the vector Ward identities are recovered as a consequence of the low
energy limit fixing. The axial Ward identity must be accepted as violated.
This is a consequence of an arbitrary choice. Note that this point of view
for the divergences gives a fundamental character to the arbitrariness
involved in the triangle anomalies. The arbitrariness involved in the
redefinition of the calculated expression is authorized by the
impossibilities contained in the context of the Sutherland-Veltman paradox.
Finally, it is easy to verify that the same values for the Feynman integrals
involved in the explicit evaluation of the $AVV$ (and $AAA$) anomalous
amplitude will lead us to symmetry preserving results for all other
three-point functions.

After these important remarks the conclusion of the present investigation
can be safely stated: If we define a strategy to handle the divergences
through the consistency conditions (i.e., if we adopt a regularization
belonging to what we denominate the first class) all amplitudes will become
nonambiguous and symmetry preserving where they must be, and they will
exhibit the correct violations where they need to. The violations naturally
appear with the correct value required by the phenomenology, but without
having to admit the ambiguous character. The final picture that has emerged
from this analysis is in accordance with our expectations. If there are
quantum physical phenomena in nature, as the anomalies seem to be, it is
desirable to justify them without resorting to typical ingredients of
perturbative solutions, such as ambiguities or infinities. This is due to
the fact that in an exact solution, which would certainly be free from
divergences and ambiguities, the anomalies must still be present. At this
point, it seems that a clean and sound conclusion can be stated. A
consistent interpretation for the perturbative amplitudes cannot admit them
as ambiguous. The ambiguities represent violations of fundamental symmetries
used to construct the theories and the destruction of the predictive power
of the QFT's. In fact, they are not necessary in any case; not even for the
justification of the triangle anomalies. Adopting this point of view, some
clarifications can be immediately obtained in many controversies present
nowadays in the literature, where the manipulation and calculation of
divergent amplitudes play a crucial role \cite{NJL}\cite{ORIMAR-PRD}\cite
{TENSOR}\cite{CPT1}\cite{CPT2}.

{\bf Acknowledgments}: G.D. acknowledges a grant from CNPq/Brazil and O.A.B.
from FAPERGS/Brazil.

\end{document}